\documentclass{emulateapj}

\usepackage{amssymb,amsmath}
\usepackage{graphicx}
\usepackage{natbib}
\usepackage{aas_macros}

\begin{document}

\title{Impact of supernova and cosmic-ray driving on the surface brightness of the galactic halo in soft X-rays}

\author{Thomas Peters\altaffilmark{1}}
\email{tpeters@mpa-garching.mpg.de}
\author{Philipp Girichidis\altaffilmark{1}}
\author{Andrea Gatto\altaffilmark{1}}
\author{Thorsten Naab\altaffilmark{1}}
\author{Stefanie Walch\altaffilmark{2}}
\author{Richard W\"{u}nsch\altaffilmark{3}}
\author{Simon C. O. Glover\altaffilmark{4}}
\author{Paul C. Clark\altaffilmark{5}}
\author{Ralf S. Klessen\altaffilmark{4}}
\author{Christian Baczynski\altaffilmark{4}}
  
\altaffiltext{1}{Max-Planck-Institut f\"{u}r Astrophysik, Karl-Schwarzschild-Str. 1, D-85748 Garching, Germany}
\altaffiltext{2}{Physikalisches Institut, Universit\"{a}t zu K\"{o}ln, Z\"{u}lpicher Str. 77, 50937 K\"{o}ln, Germany}
\altaffiltext{3}{Astronomical Institute, Academy of Sciences of the Czech Republic, Bocni II 1401, 141 31 Prague, Czech Republic}
\altaffiltext{4}{Universit\"{a}t Heidelberg, Zentrum f\"{u}r Astronomie, Institut f\"{u}r Theoretische Astrophysik, Albert-Ueberle-Str. 2, 69120 Heidelberg, Germany}
\altaffiltext{5}{School of Physics \& Astronomy, Cardiff University, 5 The Parade, Cardiff CF24 3AA, Wales, UK}

\begin{abstract}
The halo of the Milky Way contains a hot plasma with a surface brightness in soft X-rays
of the order $10^{-12}$erg cm$^{-2}$ s$^{-1}$ deg$^{-2}$. The origin of this gas is unclear, but so far numerical models of
galactic star formation have failed to reproduce such a large surface brightness by several orders of magnitude.
In this paper, we analyze simulations of the turbulent, magnetized, multi-phase interstellar medium
including thermal feedback by supernova explosions as well as cosmic-ray feedback.
We include a time-dependent chemical network,
self-shielding by gas and dust, and self-gravity.
Pure thermal feedback alone is sufficient to produce the observed surface brightness,
although it is very sensitive to the supernova rate. Cosmic rays suppress this sensitivity and reduce the
surface brightness because they drive cooler outflows. Self-gravity has by far the largest effect
because it accumulates the diffuse gas in the disk in dense clumps and filaments, so that supernovae exploding in voids can eject a large amount of
hot gas into the halo. This can boost the surface brightness by several orders
of magnitude. Although our simulations do not reach a steady state, all simulations produce surface brightness values of
the same order of magnitude as the observations, with the exact value depending sensitively on the
simulation parameters.
We conclude that star formation feedback alone
is sufficient to explain the origin of the hot halo gas, but measurements
of the surface brightness alone do not provide useful diagnostics for the study of galactic star formation.
\end{abstract}

\maketitle

\section{Introduction}

It has long been known that the halo of the Milky Way contains a hot, diffuse plasma that can be observed in soft X-rays
\citep[e.g.][]{burmen91,wangyu95,pietzetal98,snowdenetal98,kunsno00,smithetal07,henshe10}. Most recently, \citet{henshe13}
obtained {\em XMM-Newton} observations at energies of $0.5$--$2.0\,$keV for 110 sightlines and inferred the X-ray
temperature and surface brightness. They found the surface brightness to vary by one order of magnitude
$\sim 0.5$--$7 \times 10^{-12}$erg cm$^{-2}$ s$^{-1}$ deg$^{-2}$, while the temperature had a rather constant value
around $\sim 2 \times 10^6\,$K.

What produces this hot halo gas is currently not clear. Two main processes have been dicussed, namely
winds driven by star formation feedback from the Galactic disk \citep[e.g.][]{shafie76,bregman80,norike89,joungmaclow06,hilletal12}
and galactic accretion flows \citep[e.g.][]{toftetal02,rasmussenetal09,crainetal10}. Other ideas invoke
supernova explosions more than $100\,$pc above the galactic disk \citep{shelton06,henshe09}
or hybrid cosmic-ray and thermally driven outflows \citep{ever08}. However, the relative importance
of these different potential sources of the hot halo gas is unknown.

Recently, \citet{henleyetal15} analyzed stratified
box simulations of a supernova-driven galactic disk by \citet{hilletal12} and measured surface brightnesses that were two
orders of magnitude or more too low compared to observations. The origin of this discrepancy is unclear, although
\citet{henleyetal15} discuss several possibilities.

In this Letter, we study the galactic halo X-ray emission of simulations from the SILCC project\footnote{SImulating the Life-Cycle of molecular Clouds,
http://hera.ph1.uni-koeln.de/$\sim$silcc/} \citep{walchetal15,girisilcc}. We model a non-equilibrium multi-phase interstellar medium using a time-dependent
chemical network and include feedback by supernovae and cosmic rays
(Section~\ref{sec:sim}).
In Section~\ref{sec:prof} we show vertical profiles of the galactic winds.
We create synthetic surface brightness maps (Section~\ref{sec:surf}) of the galactic halo and study the
morphology (Section~\ref{sec:morph}), surface brightness distribution (Section~\ref{sec:dist}) and time evolution (Section~\ref{sec:time})
of the X-ray emission.
In Section~\ref{sec:ebud} we discuss the energy budget of the halo gas. We conclude in Section~\ref{sec:con}.

\section{Simulations}
\label{sec:sim}

We analyze simulations run with the \texttt{FLASH}~4.1 adaptive mesh refinement code \citep{fryxell00,dubey09},
employing a positivity-preserving magnetohydrodynamics solver \citep{bouchut07,waagan09}.
The magnetohydrodynamic
equations are extended by a mono-energetic, relativistic cosmic-ray fluid
\citep{yangetal12} in the advection-diffusion approximation as described in \citet{girietal14}.
Parallel to the magnetic field, the diffusion tensor is
$10^{28}\,\mathrm{cm^2\,s^{-1}}$, and in perpendicular direction $10^{26}\,\mathrm{cm^2\,s^{-1}}$.

All simulations use a time-dependent chemical network \citep{nellan97,glovmcl07,glovmcl07b,glovetal10,glovclar12},
which follows the abundances of free electrons, H$^{+}$, H, H$_{2}$, C$^{+}$, O and CO. Dust shielding and molecular self-shielding
are taken into account with the \texttt{TreeCol} algorithm (\citealt{clarketal12}, W\"{u}nsch et al. in prep).
In hot gas, we use the \cite{gnafer12} cooling rates and include
diffuse heating from the photoelectric effect, cosmic rays and X-rays
following the prescriptions of \citet{baktie94}, \citet{gollan78} and \citet{woletal95}, respectively. More information
on the chemical network and the various heating and cooling processes can be found in \citet{gattoetal15} and \citet{walchetal15}.

We simulate stratified boxes with dimensions $2\,$kpc$ \times 2\,$kpc$ \times 40\,$kpc or $1\,$kpc$ \times 1\,$kpc$ \times 40\,$kpc
and a grid resolution of $15.6\,$pc. These boxes are larger and have a coarser resolution than previous simulations of similar type
\citep{walchetal15,girisilcc} because our cosmic ray diffusion coefficient does not allow for higher resolution.
The simulation domain is periodic in the $x$ and $y$ directions. Gas can leave the box in $z$ direction, but infall is not permitted.

We use a tree-based method to incorporate self-gravity (W\"{u}nsch et al. in prep.) and additionally impose an external
gravitational potential for the stellar component of the gravitational force on the gas. We choose
the parameters of the external potential to fit solar neighbourhood values, namely a surface density in stars of $\Sigma_{*} = 30\,M_\odot\,$pc$^{-2}$
and a vertical scale height of $z_\mathrm{d} = 100\,$pc.

The disk has a gas surface density $\Sigma_\mathrm{gas} = 10\,M_\odot\,$pc$^{-2}$. The mass is initially set up according
to a Gaussian distribution in $z$-direction with a scale height of $60\,$pc.
The disk has a midplane density of $\rho_\mathrm{m} = 8.7 \times 10^{-24}\,$g\,cm$^{-3}$ and temperature of $T_\mathrm{m} = 4700\,$K.
It is embedded in a halo with density $\rho_\mathrm{h} = 10^{-28}\,$g\,cm$^{-3}$ and temperature $T_\mathrm{h} = 4.1 \times 10^8\,$K.
Magnetic fields are initialized in the $x$-direction and scale in magnitude
with the square root of the gas density. The midplane field strength is $B_\mathrm{m} = 9.8\,$nG.
For more information on the initial conditions and simulation setup see \citet{walchetal15} and \citet{girisilcc}.

Following the Kennicutt-Schmidt relation \citep{schmidt59,kenn98}, we inject supernova explosions at a constant supernova rate surface density
of $60\,$Myr$^{-1}\,$kpc$^{-2}$ and half of this value (see discussion in \citealt{walchetal15}).
Using the same approach as \citet{avilbreit04} and \citet{joungmaclow06},
explosion events are drawn randomly from a distribution of type Ia and type II supernovae, taking
spatial and temporal correlations of explosion sites in stellar clusters into account (see \citealt{girisilcc}).

With each supernova explosion, we inject a thermal energy of $E_\mathrm{SN} = 10^{51}\,$erg.
For a detailed description of the supernova injection subgrid model we refer
to \citet{gattoetal15}.
In simulations with cosmic rays, we additionally
inject 10\% of $E_\mathrm{SN}$ in the form of cosmic ray energy. We also consider a control run in which we only inject cosmic ray energy,
but no thermal energy during supernova explosions \citep{giricr}.

Table~\ref{tab:sims} summarizes the key simulation properties for the runs presented in this paper. Simulations SN, SNCR and CR 
are discussed in \citet{giricr}. For the small boxes, we also present simulations that include self-gravity, and simulations with half the fiducial supernova rate.
Additionally, we show one high-resolution simulation from \citet{walchetal15} with identical supernova rate and surface density, and which uses the same clustered driving, S10-KS-clus.

\begin{table*} 
\caption{Key simulation parameters.}
\label{tab:sims}
\begin{center}
\begin{tabular}{lcccccccc}
\hline
\hline
simulation      & $\Sigma_\mathrm{gas}$  & supernova rate          & resolution & box dimension  & thermal  & cosmic & self- \\
name            & ($M_\odot\,$pc$^{-2}$) & (Myr$^{-1}\,$kpc$^{-2}$) & (pc)       & (kpc$^{3}$)    & feedback & rays & gravity \\
\hline
SN            & $10$ & $60$ & $15.6$ & $2 \times 2 \times 40$  & yes & no  & no\\
SNCR            & $10$ & $60$ & $15.6$ & $2 \times 2 \times 40$  & yes & yes & no\\
CR            & $10$ & $60$ & $15.6$ & $2 \times 2 \times 40$  & no  & yes & no\\
\hline
sb-SN         & $10$ & $60$ & $15.6$ & $1 \times 1 \times 40$ & yes & no  & no\\
sb-SN-lo      & $10$ & $30$ & $15.6$ & $1 \times 1 \times 40$ & yes & no  & no\\
sb-SN-sg      & $10$ & $60$ & $15.6$ & $1 \times 1 \times 40$  & yes & no  & yes\\
sb-SNCR         & $10$ & $60$ & $15.6$ & $1 \times 1 \times 40$  & yes & yes & no\\
sb-SNCR-lo      & $10$ & $30$ & $15.6$ & $1 \times 1 \times 40$  & yes & yes & no\\
sb-SNCR-sg      & $10$ & $60$ & $15.6$ & $1 \times 1 \times 40$  & yes & yes & yes\\
\hline
S10-KS-clus     & $10$ & $60$ & $3.9$ & $0.5 \times 0.5 \times 10$  & yes & no & yes\\
\hline
\end{tabular}
\medskip\\
We list the values of the gas surface density $\Sigma_\mathrm{gas}$, the supernova rate (surface density), the numerical resolution,
and the box dimension. We also specify which
simulations include thermal energy injection, cosmic rays or self-gravity.
\end{center}
\end{table*}

\section{Vertical profiles}
\label{sec:prof}

\begin{figure*}
\centerline{\includegraphics[height=160pt]{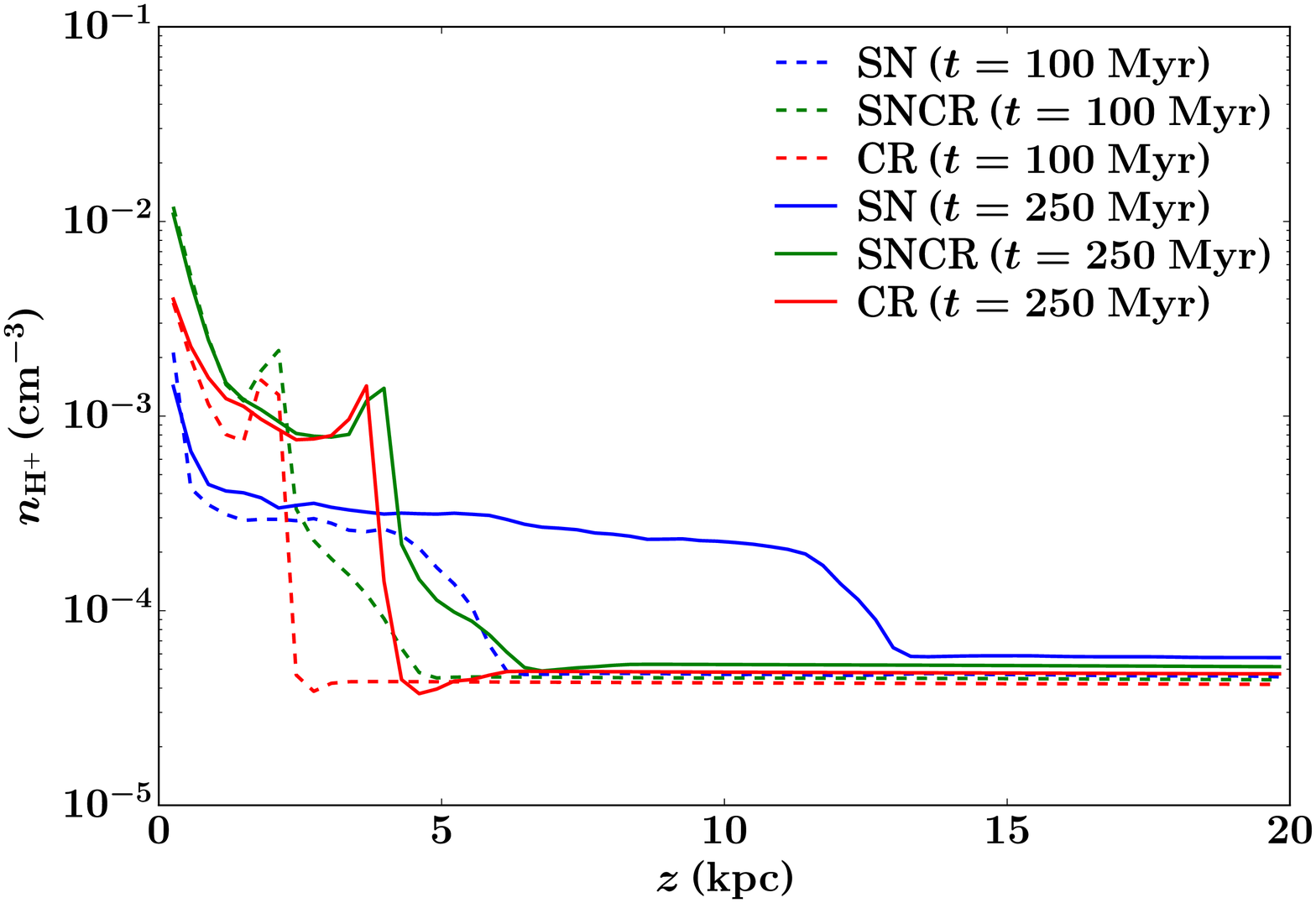}
\includegraphics[height=160pt]{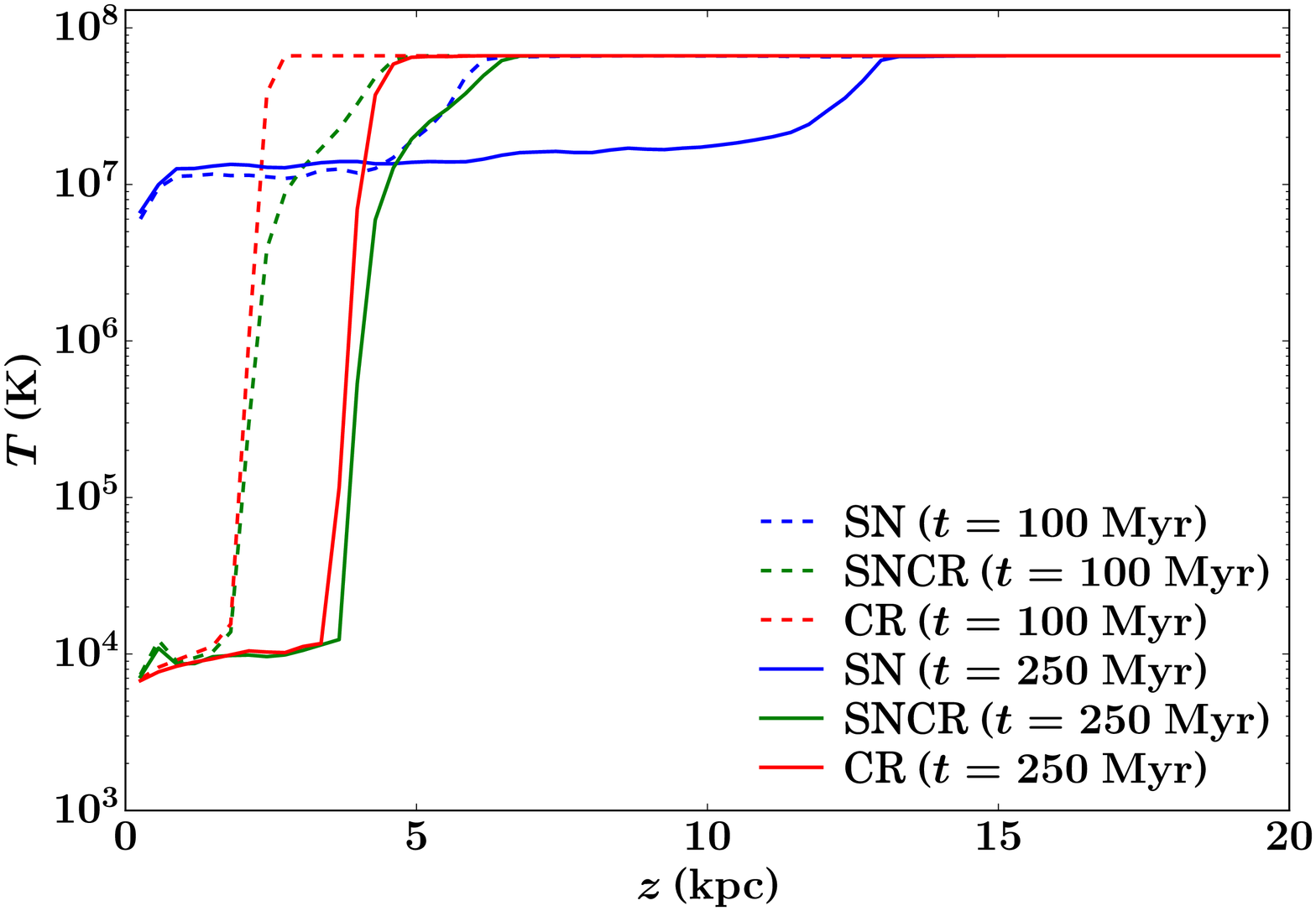}}
\caption{H$^+$ number density and temperature profiles of the simulations SN, SNCR and CR after $100\,$Myr (dashed) and $250\,$Myr (solid).}
\label{fig:prof}
\end{figure*}

Since X-rays are emitted from hot, ionized gas in the halo, we show in 
Figure~\ref{fig:prof} vertical profiles of the H$^+$ number density and temperature for the upper halo
of the simulations SN, SNCR and CR after $100$ and $250\,$Myr. The simulations with cosmic-ray feedback
can reach half an order of magnitude higher densities of H$^+$ in the halo.
However, this H$^+$ is at much lower temperatures, $T \approx 10^4\,$K instead of $T \approx 10^7\,$K
in the pure thermal case. Including cosmic rays, the galactic wind is mostly
atomic in composition \citep{giricr}, which through mixing processes with the ionized component leads to a lower
average temperature of H$^+$ in the halo.

After $250\,$Myr, simulation SN has filled the halo up to 13\,kpc with
hot gas from a galactic wind. Runs SNCR and CR expand more slowly (see discussion in \citealt{giricr}). 
Simulation SNCR only drives gas in excess of $10^4\,$K into the halo in a $\sim 2\,$kpc
wide layer ahead of the primarily atomic outflow, while run CR does not create any hot gas at all.

\section{Surface brightness maps}
\label{sec:surf}

To estimate the importance of absorption for the soft X-ray surface brightness maps, we compute the total optical depth for the neutral hydrogen
cross section along lines of sight from $0.1\,$kpc above the galactic midplane to the end of the simulation box at 20\,kpc.
Soft X-rays are primarily absorbed by atomic hydrogen near $0.5\,\mathrm{keV}$. The absorption cross section per hydrogen nucleus can be
approximated by
$\sigma(E) = 2.27 \times 10^{-22}\,\mathrm{cm}^2 (E /\,\mathrm{keV})^{-2.485}$, where $E$ is the energy of the X-ray photon \citep{shangetal02}.
The function $\sigma(E)$ falls monotonically from $\sigma(0.5\,\mathrm{keV}) = 1.27 \times 10^{-21}\,\mathrm{cm}^2$ to
$\sigma(2.0\,\mathrm{keV}) = 4.05 \times 10^{-23}\,\mathrm{cm}^2$,
with an average value of $\overline{\sigma} = 2.49 \times 10^{-22}\,\mathrm{cm}^2$.

We find optical depths well below $0.1$ for the vast majority of the sightlines.
Although the galactic wind is mostly atomic in composition in the cosmic-ray runs, the total
amount of atomic hydrogen that accumulates in the halo remains small, and
the entire halo stays optically thin. Therefore, we can neglect absorption for the computation of the surface brightness maps.

We use an emission table generated with the Astrophysical Plasma Emission Code (APEC, \citealt{smithetal01})
from the collisional ionization database AtomDB\footnote{http://www.atomdb.org} to compute the X-ray emissivity in the energy band $0.5$--$2.0\,$keV.
The metallicity is solar throughout. The emissivity is largest for gas with temperatures between $10^6\,$K and $10^7\,$K and declines
rapidly for higher and lower temperatures.

Given the emissivity as function of temperature and density, we compute surface brightness maps by integrating the emissivity
along lines of sight. The base grid of our simulations has a resolution of $128^2$ in the midplane ($x$-$y$ plane) for the large
boxes and $64^2$ for the small ones. For each
of these $(x,y)$-coordinates, we trace a ray outwards until the end of the simulation volume and integrate the emissivity along the ray.
We start the integration at an altitude of $z = 0.1$\,kpc above the midplane to only include halo gas.
Hereafter, we call the region $z \geq 0.1$\,kpc the upper halo.
To visualize the structure of the galactic wind in the simulations, we also generate edge-on maps of the surface
brightness with an analogous procedure.

\citet{henleyetal15} use a more complex method to post-process the galactic fountain models of \citet{hilletal12}, following
closely the analyis of real observations.
We have compared our values for the surface brightness to the results of this more sophisticated procedure
by applying our method to the \citet{hilletal12} simulations. We find
good agreement in the surface brightness distribution between the results reported by \citet{henleyetal15}
and our own values. However, our approach does not allow us to determine X-ray temperatures, and so we will focus exclusively
on the surface brightness in this paper.

\section{Morphology of the X-ray emission}
\label{sec:morph}

\begin{figure*}
\centerline{\includegraphics[height=90pt]{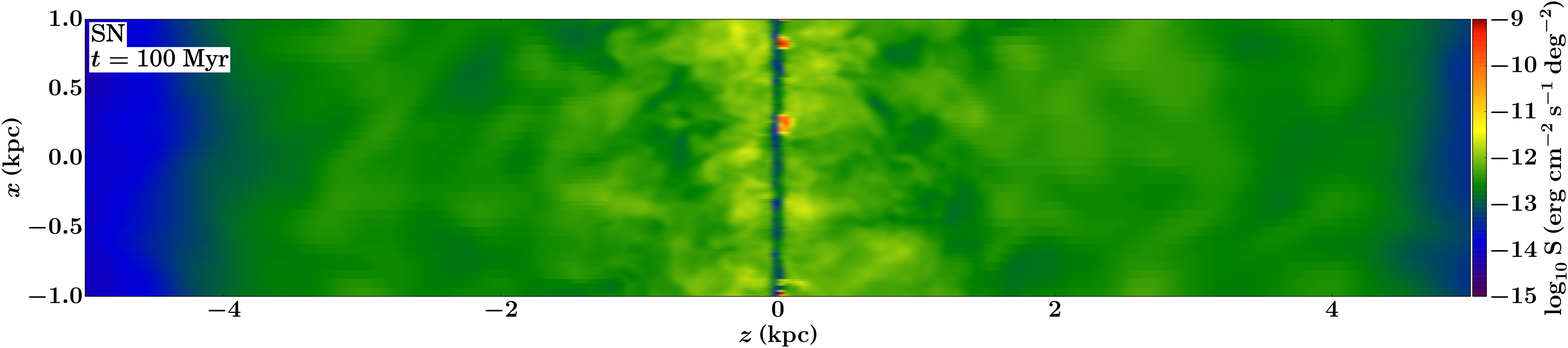}
\includegraphics[height=90pt]{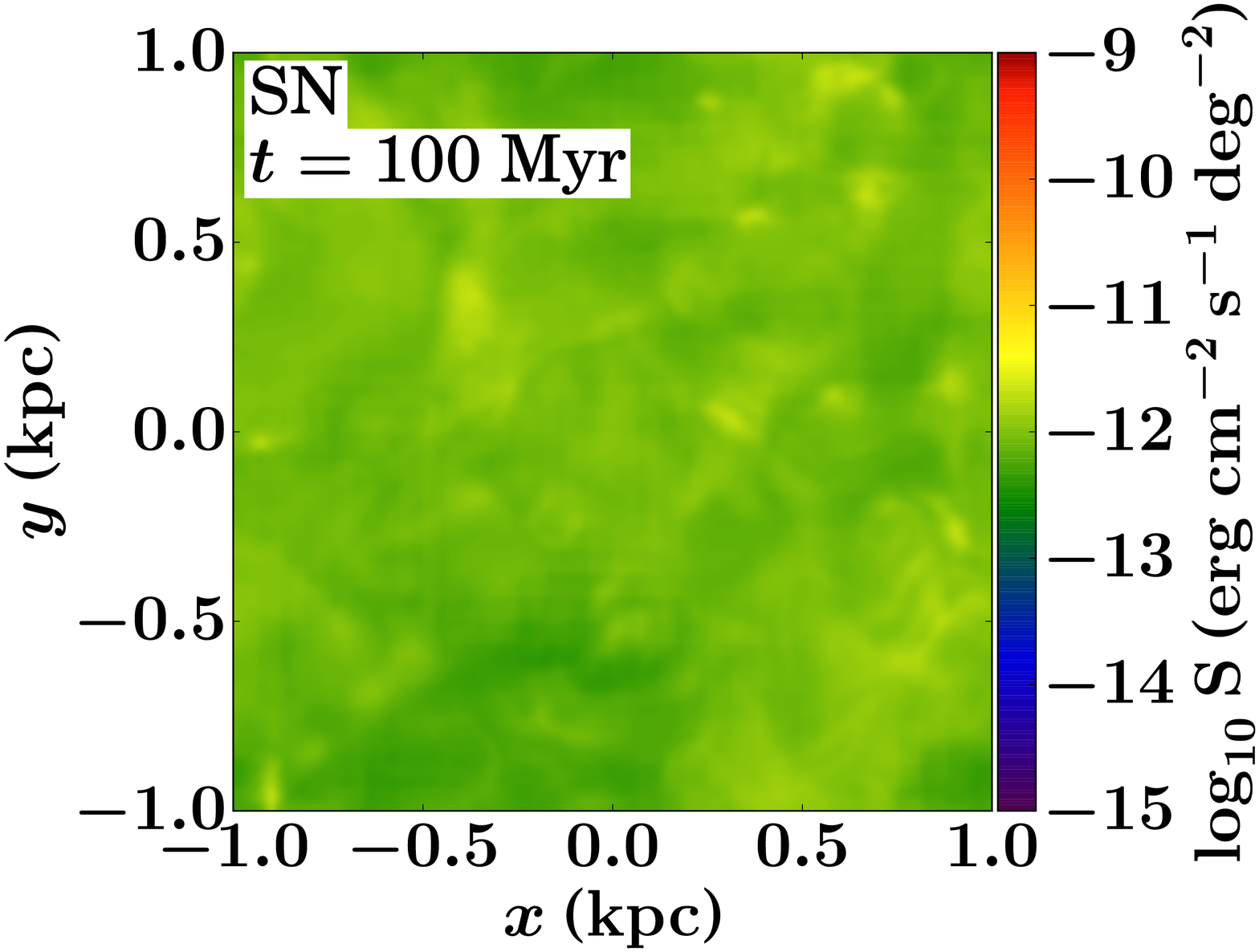}}
\centerline{\includegraphics[height=90pt]{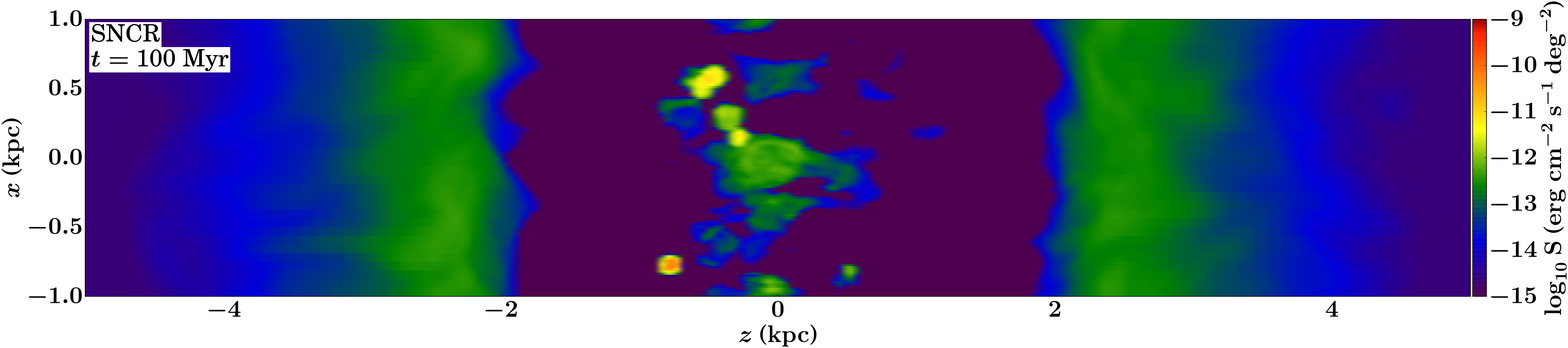}
\includegraphics[height=90pt]{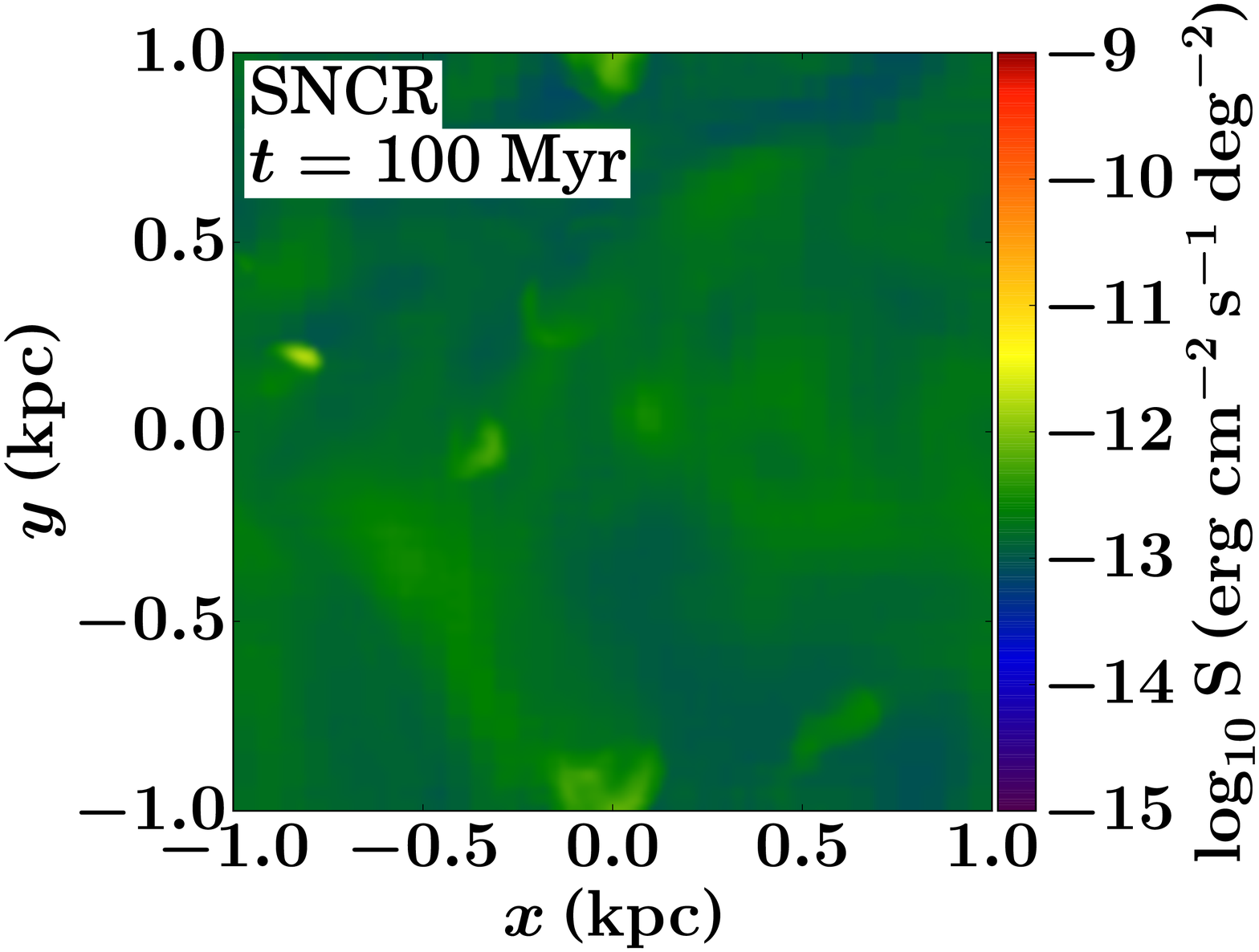}}
\centerline{\includegraphics[height=90pt]{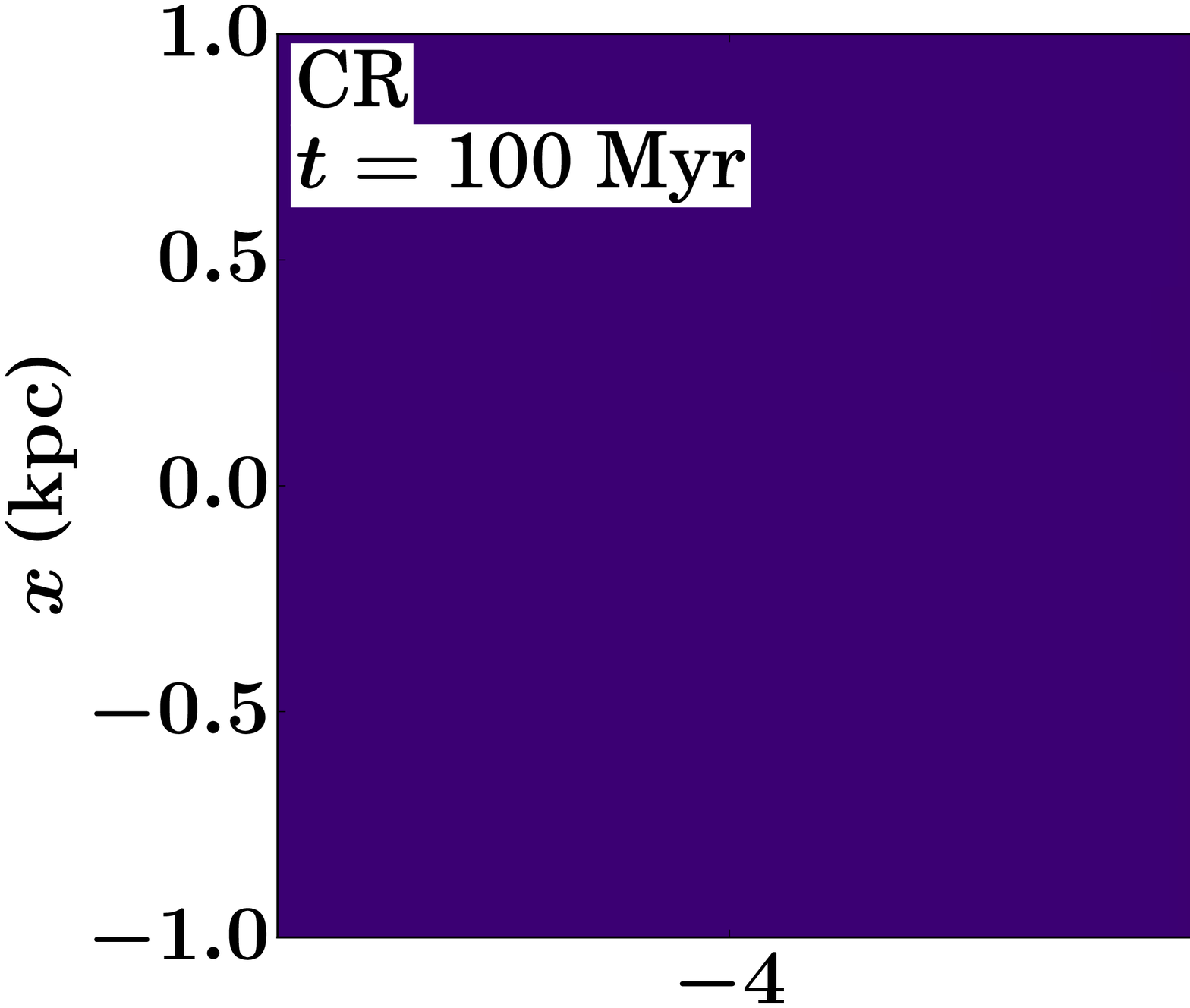}
\includegraphics[height=90pt]{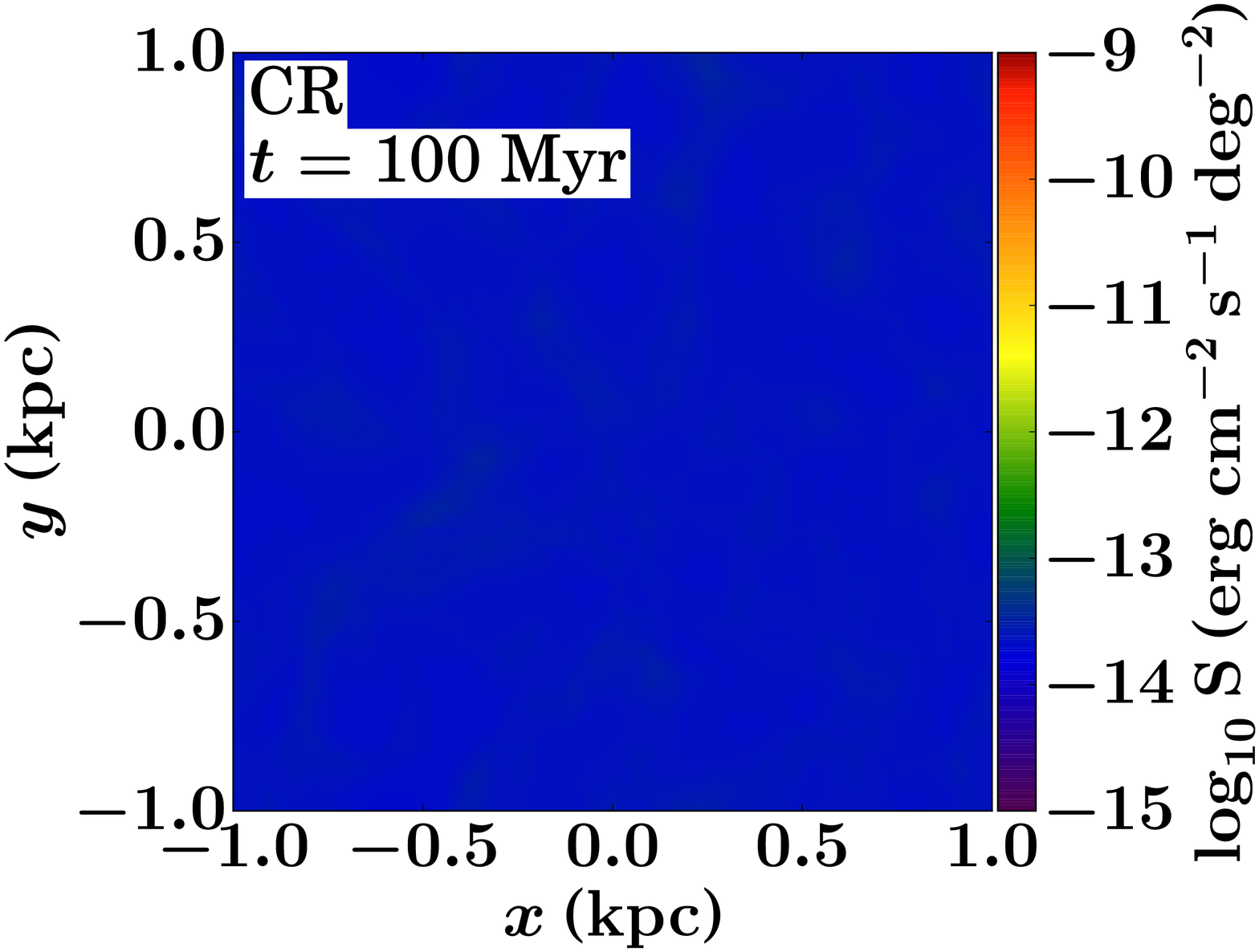}}
\caption{Edge-on (left) and face-on (right) surface brightness in soft X-rays (0.5--2.0 keV) in units of erg cm$^{-2}$ s$^{-1}$ deg$^{-2}$ for simulation
(from top to bottom) SN, SNCR and CR at 100 Myr.}
\label{fig:morph}
\end{figure*}

The morphology of the face-on soft X-ray emission is qualitatively similar in all simulations. The largest differences occur as function
of the outflow driving mechanism. We illustrate this in Figure~\ref{fig:morph} for the three simulations SN, SNCR and CR at a simulation time
of $t = 100\,$Myr. When thermal feedback is included, the supernova and/or cosmic-ray driving launches a wind into the galactic halo that is bright in
soft X-rays. In the case of runs SN and SNCR, supernova explosions inject hot gas into the halo that reaches a height of $4\,$kpc.
The edge-on surface brightness corresponds very well to the vertical profiles of Figure~\ref{fig:prof}.

Because of the lower temperature of the wind, run SNCR has a smaller face-on surface brightness than run SN.
Most of the gas launched by the galactic wind in run SNCR is too cool to emit soft X-rays.
Therefore, supernovae which explode at high altitudes and inject thermal energy locally into the halo provide a more important contribution to the face-on surface brightness
than in run SN, where the galactic wind is bright in soft X-rays and the emission stems from the entirety of the halo gas.
As expected, run CR is completely dark in soft X-rays, with the exception of the hot gas remaining from the initial conditions.
In all cases
the face-on maps of surface brightness are relatively homogeneous, with a scatter
for different lines of sight of less than an order of magnitude. Run SN is an order of magnitude brighter
than run SNCR, with run SN being very close to the observed values.

\section{Surface brightness distribution}
\label{sec:dist}

\begin{figure*}
\centerline{\includegraphics[height=160pt]{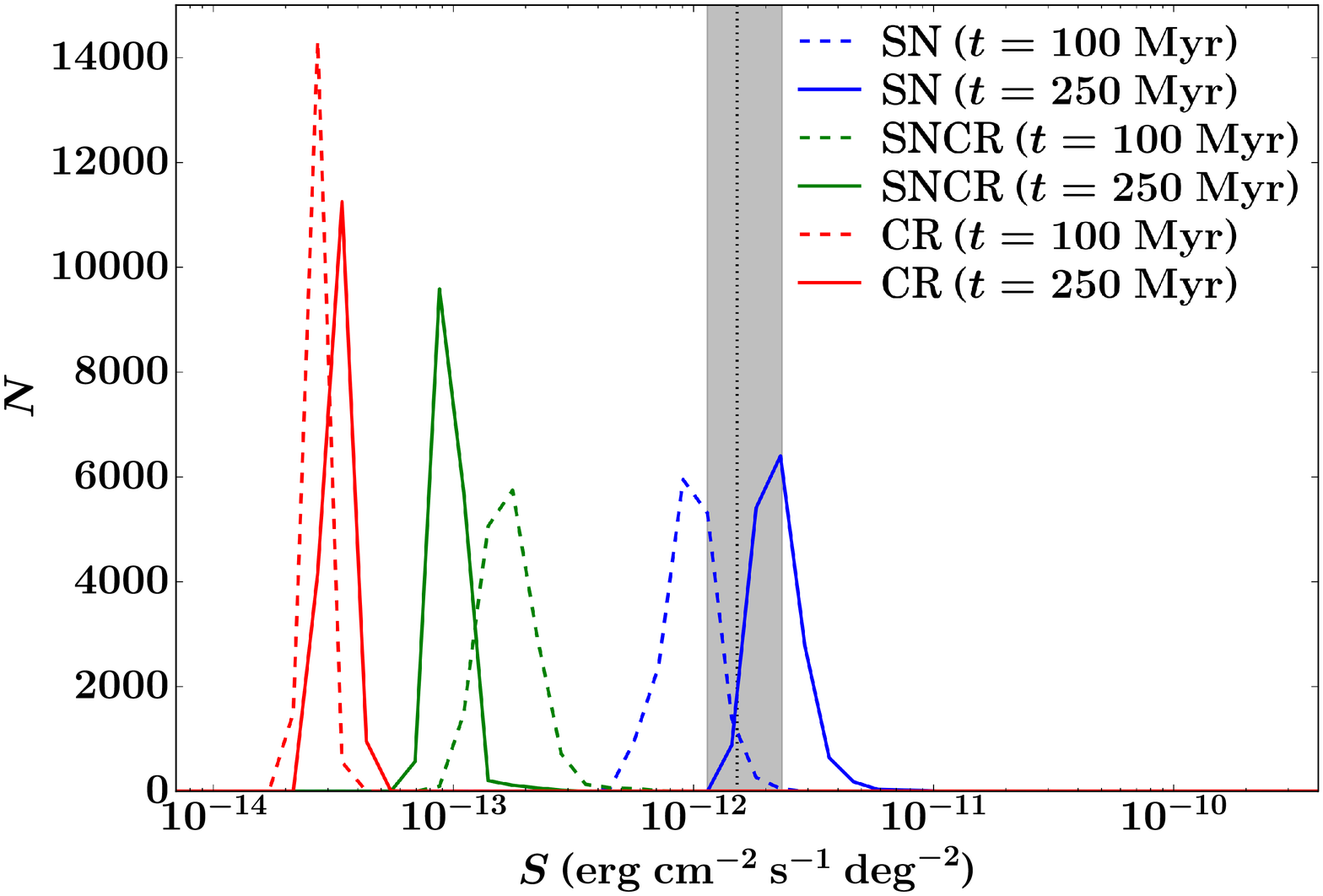}
\includegraphics[height=160pt]{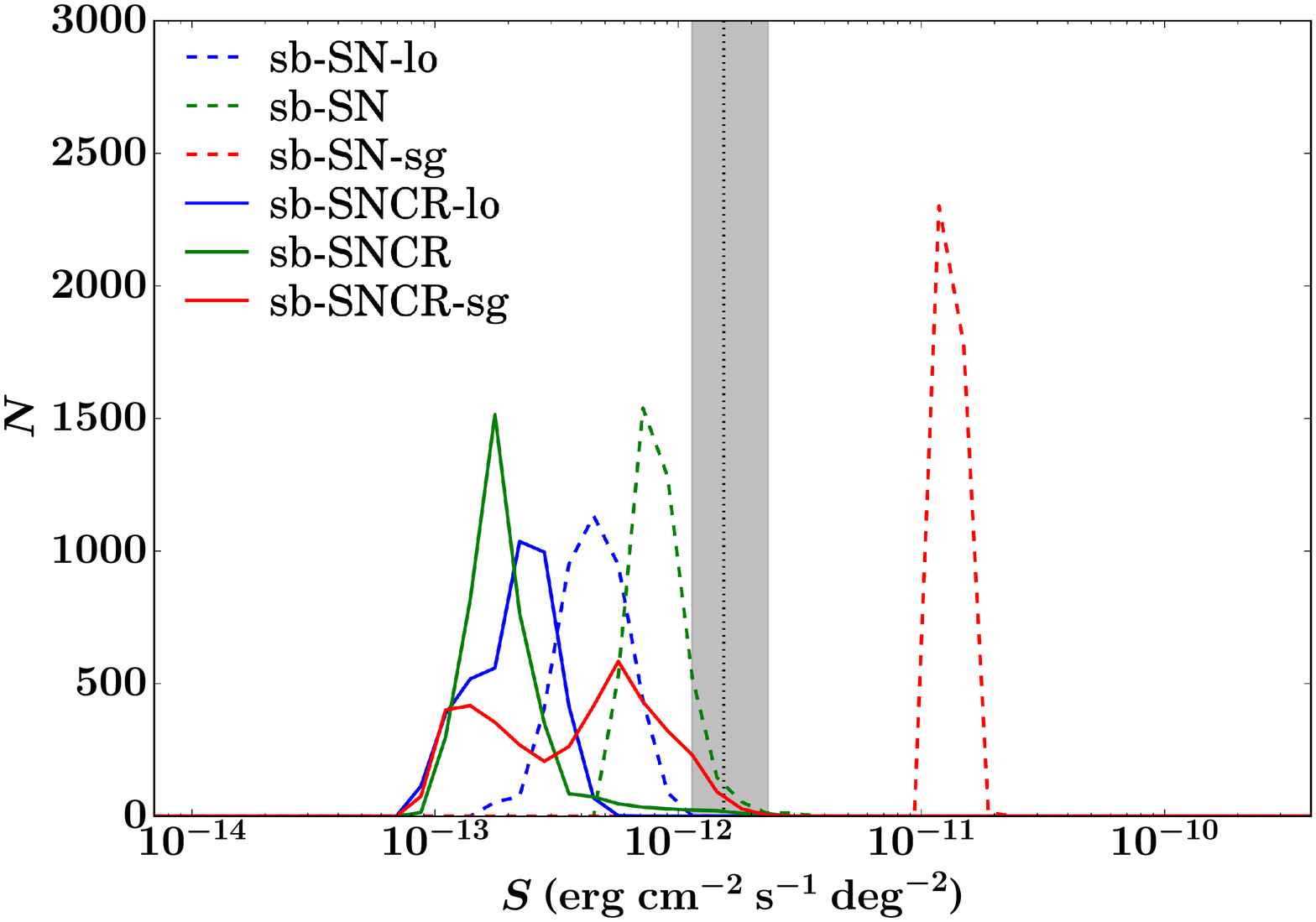}}
\caption{Histograms of the face-on surface brightness distribution. Left: Time evolution of the simulations SN, SNCR and CR. Right:
SN and SNCR runs with small simulation volumes at $t = 100\,$Myr. The dotted black line represents the median of the
\citet{henshe13} observations and the gray area the interquartile range.}
\vspace{1cm}
\label{fig:distr}
\end{figure*}

As the simulations proceed, an increasing amount of gas is pushed into the halo. Figure~\ref{fig:distr} shows the face-on surface
brightness distribution of run SN, SNCR and CR at $t = 100\,$Myr and $t = 250\,$Myr. The width of the distribution
in all snapshots is one order of magnitude or smaller. In the $150\,$Myr of evolution between the snapshots, the distribution
shifts towards larger surface brightnesses, but does not significantly change its shape.

The histograms for the simulations with the smaller volume are shown in Figure~\ref{fig:distr} for $t = 100\,$Myr.
The surface brightness distributions of the simulations with the lower supernova rates are shifted towards smaller values by a factor
of a few compared to simulations with the fiducial supernova rates, but only in the case of pure thermal feedback.
Most strikingly, the runs that include self-gravity have an up to one order of magnitude higher surface brightness than simulations
without self-gravity. Both sets of simulations with and without cosmic rays show this effect, but it is more pronounced in the pure thermal case. The reason for the increase
in surface brightness for simulations with self-gravity is the formation of filaments and sheets in the disk, which sets in after
about $50\,$Myr. These structures bind the diffuse gas in the disk, so that supernovae exploding in voids can inject large amounts
of hot gas into the halo. Without self-gravity, these ejecta are largely absorbed by diffuse gas in the disk before they can enter the halo.
All surface brightness distributions have mean values not farther away than one order of magnitude from the median of the \citet{henshe13} observations.

\section{Time evolution}
\label{sec:time}

\begin{figure*}
\centerline{\includegraphics[width=220pt]{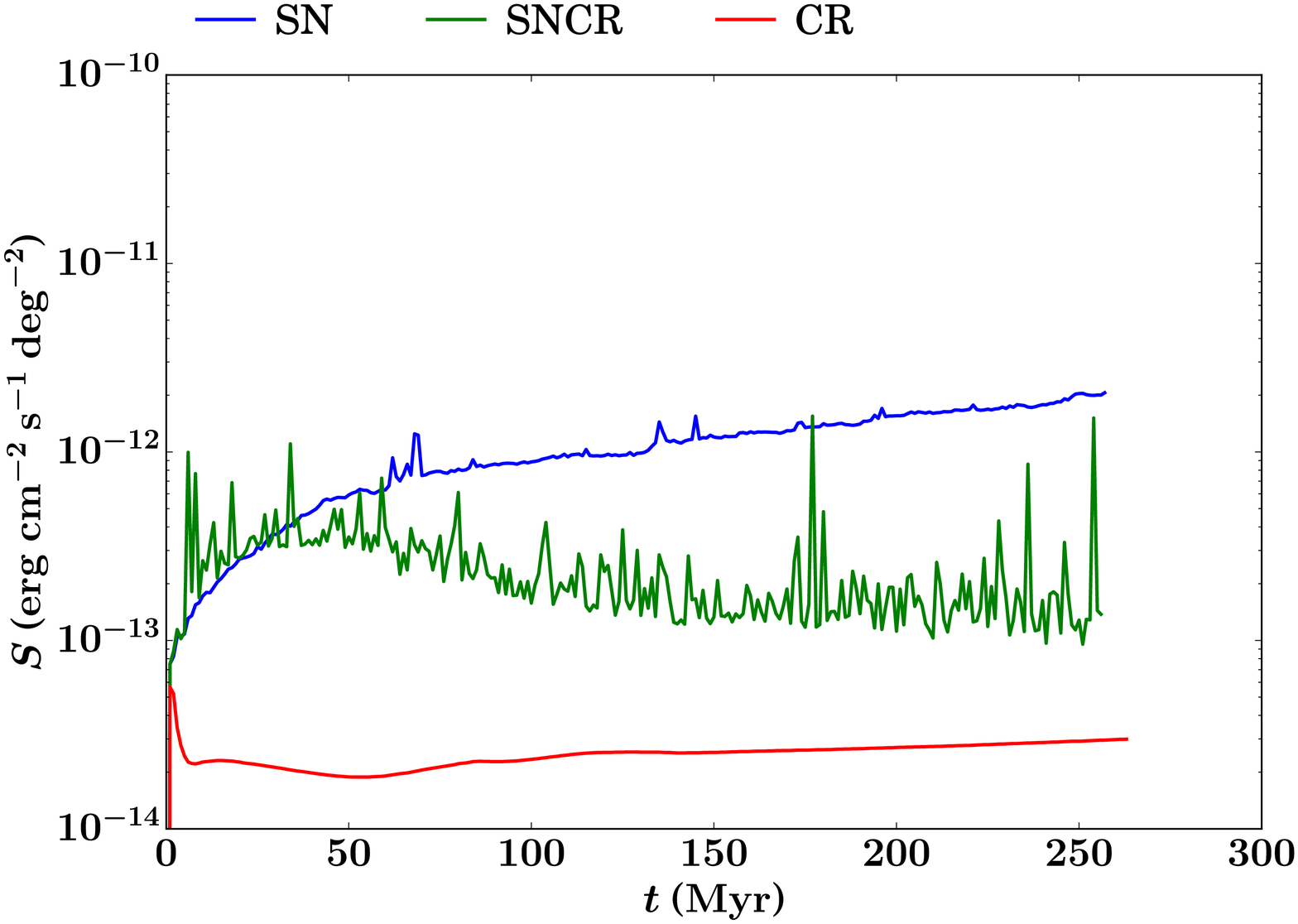}
\includegraphics[width=220pt]{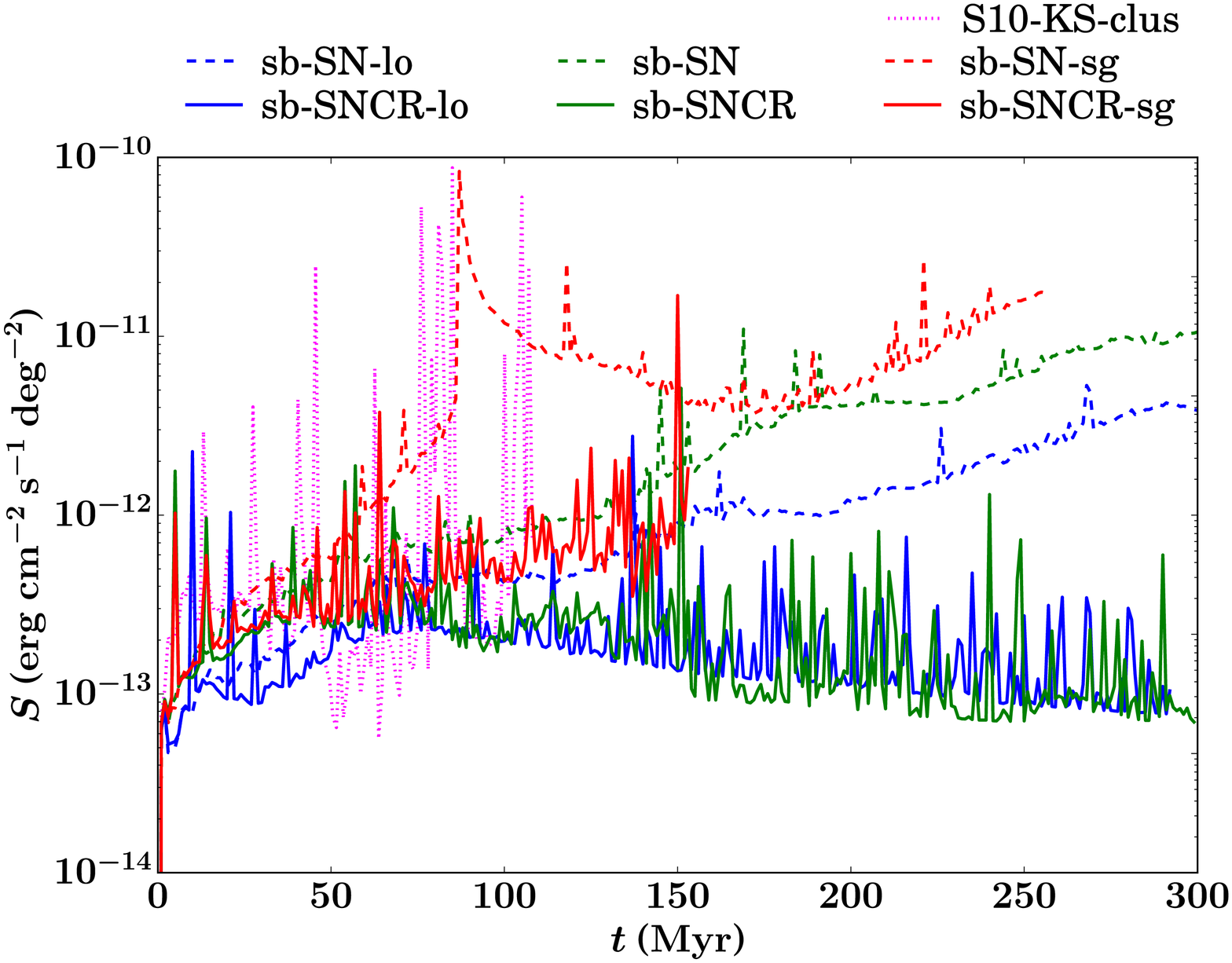}}
\centerline{\includegraphics[width=220pt]{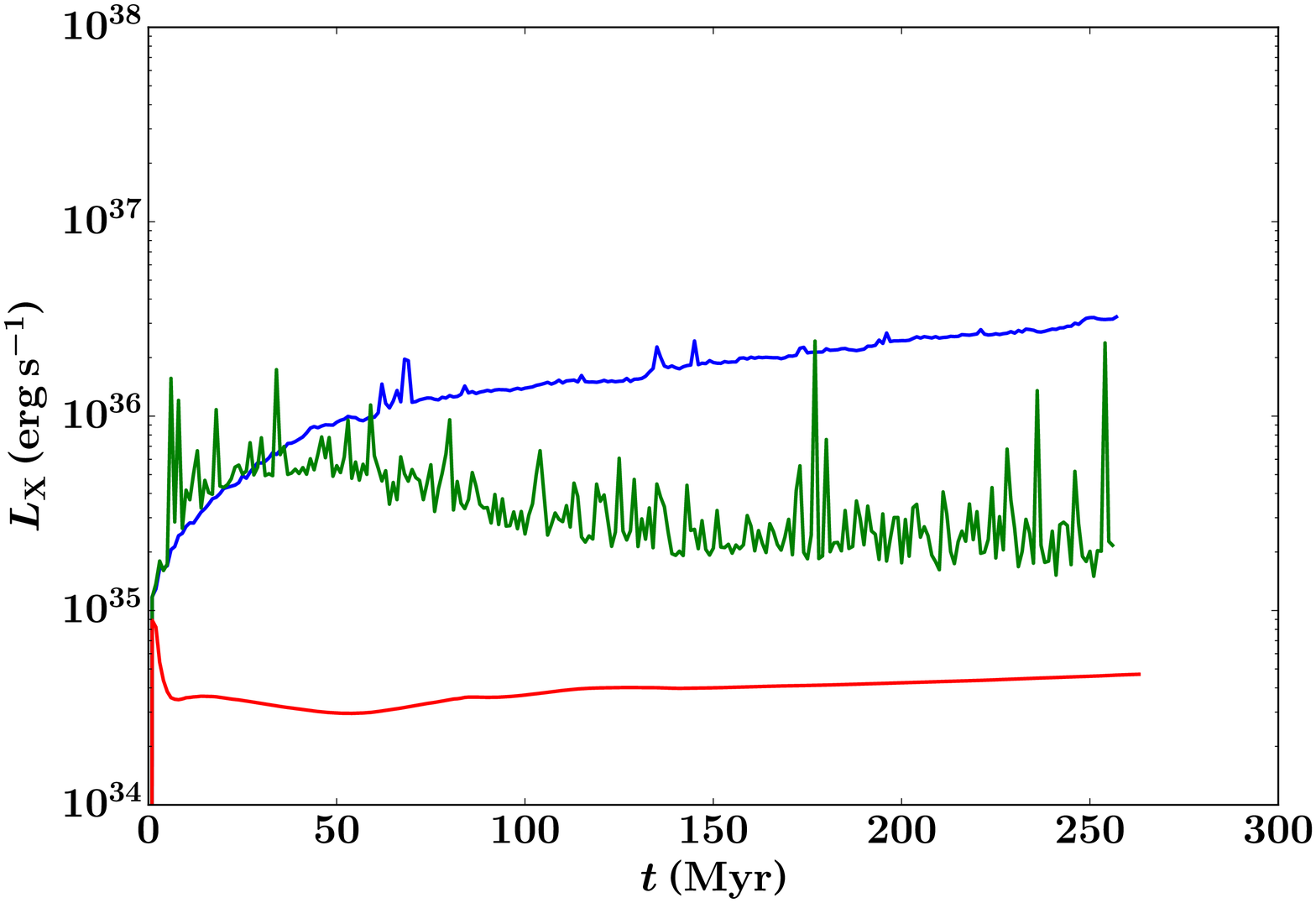}
\includegraphics[width=220pt]{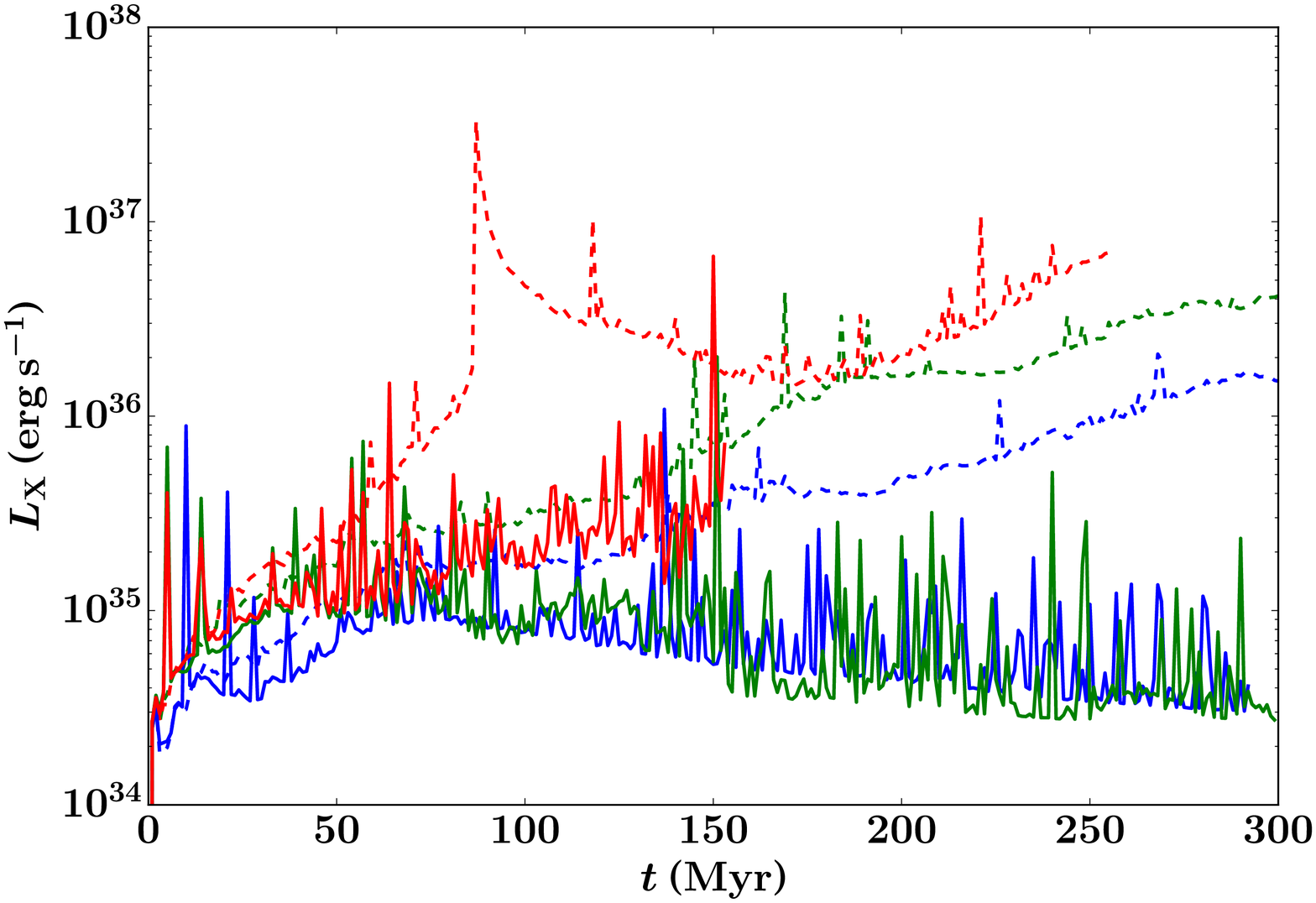}}
\centerline{\includegraphics[width=220pt]{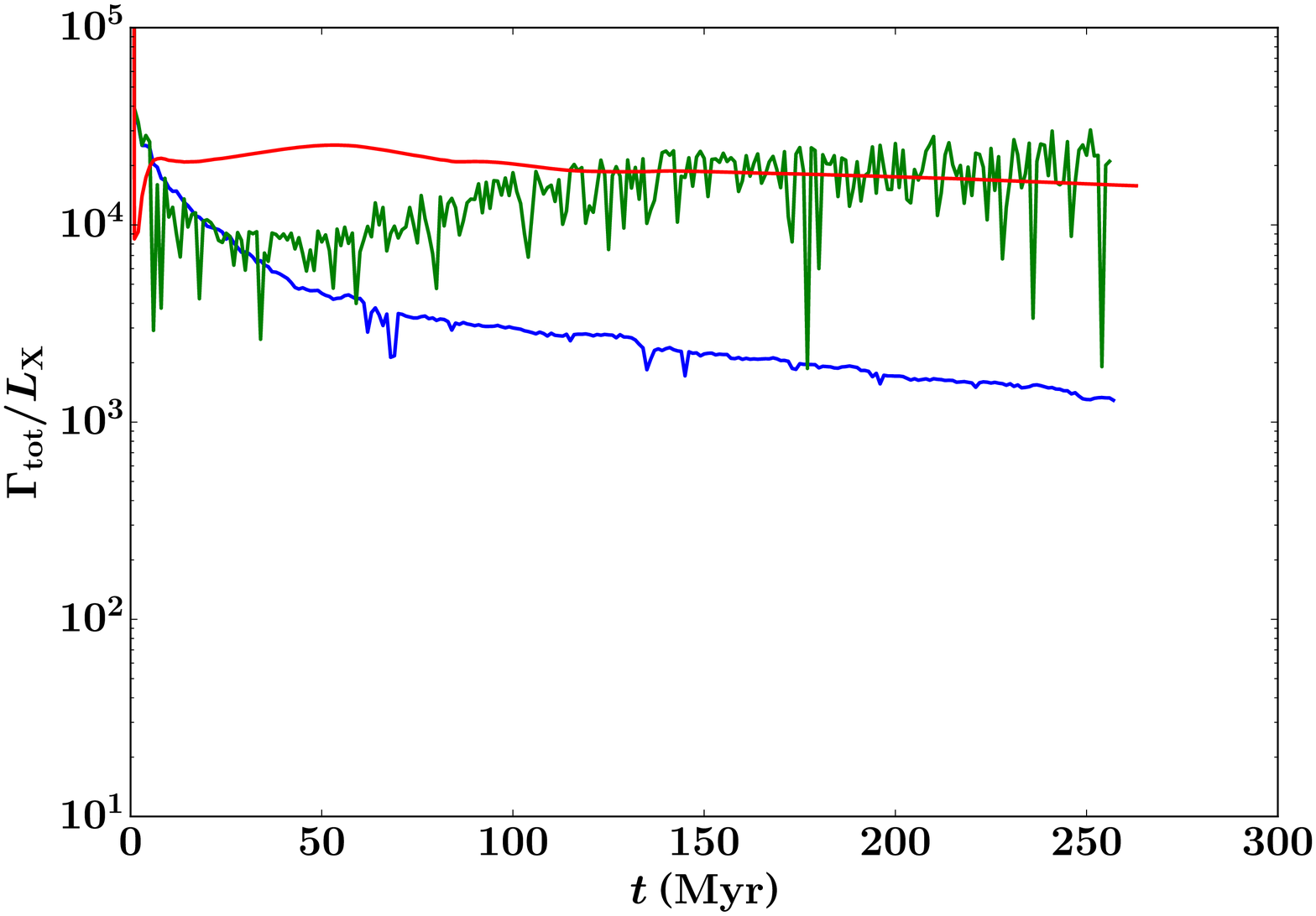}
\includegraphics[width=220pt]{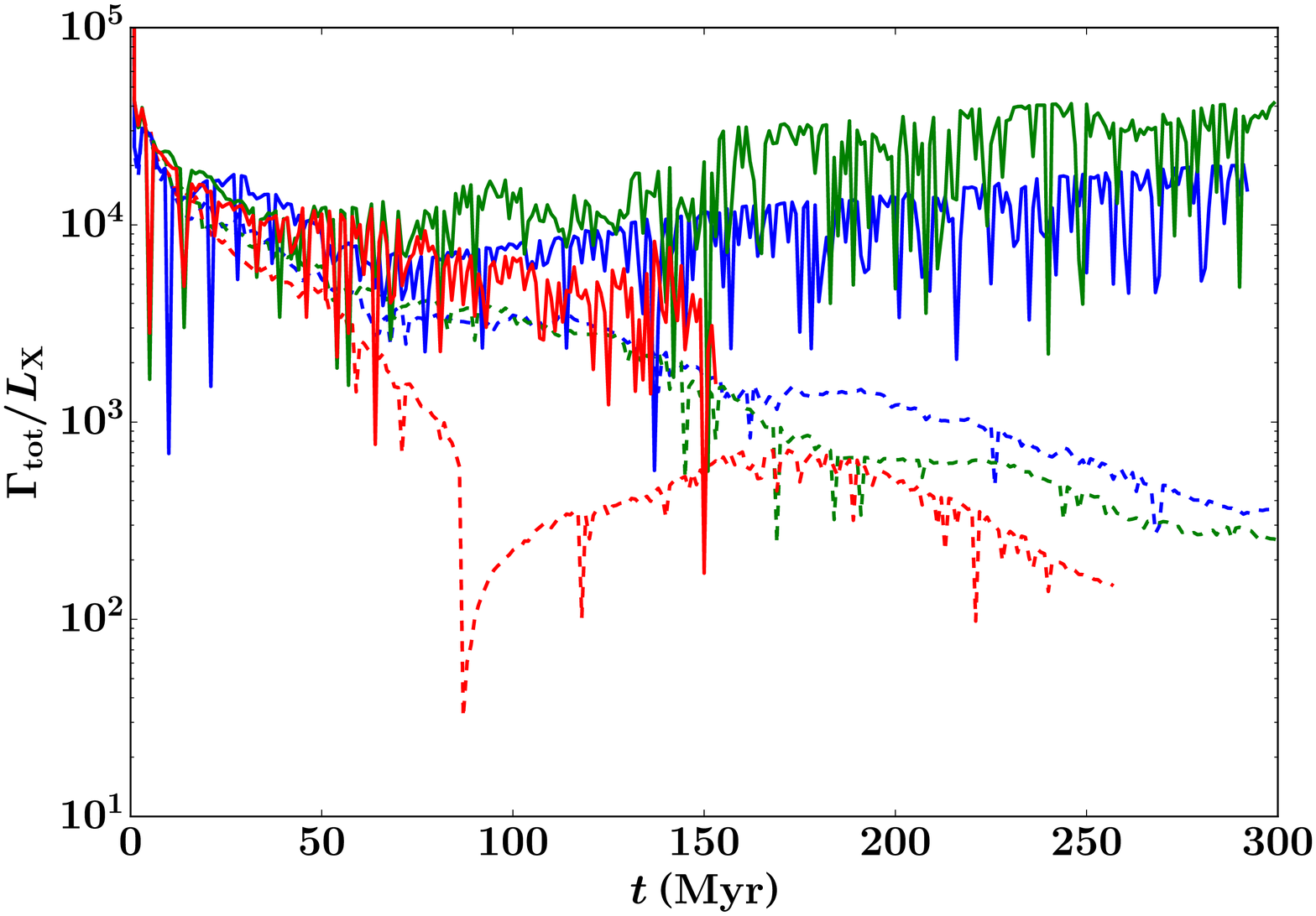}}
\caption{Mean face-on surface brightness $S$, X-ray luminosity $L_\mathrm{X}$ and ratio $\Gamma_\mathrm{tot} / L_\mathrm{X}$ for the upper halo as function of time for
the large (left) and small (right) simulation boxes.}
\label{fig:time}
\end{figure*}

The time evolution of the mean face-on surface brightness of all simulations is shown in Figure~\ref{fig:time}.
Run SN reaches a value of $S = 10^{-12}\,$erg cm$^{-2}$ s$^{-1}$ deg$^{-2}$
after 50\,Myr, and then $S$ continues to grow slowly. We note that this growth is not limited
by the size of the simulation box, which extends up to 20\,kpc (compare Figure~\ref{fig:prof}).
Run SNCR is much more intermittent than run SN because
of the large relative importance of supernova explosions at high altitudes. On average, 
the surface brightness in run SNCR is one order of magnitude smaller than in run SN.

The simulations with the smaller boxes show a similar behavior as the other runs.
For pure thermal feedback, runs with a lower supernova rate always
have a smaller surface brightness than simulations with the fiducial value.
Note, however, that the difference in surface brightness at late times is much larger than the 
difference in the supernova rate. Therefore, simulations with only thermal feedback
are very sensitive to the assumed supernova rate. When cosmic rays are included,
no systematic dependence on the supernova rate is observed.

Of all the different physical effects considered here, self-gravity has by far the largest impact.
This, however, is only a temporary
effect, because in our simulations the inclusion of self-gravity leads to a run-away process that accumulates most mass
in dense clouds that cannot be destroyed by supernovae \citep{girisilcc}. At $t = 100\,$Myr, most of the disk in run sb-SN-sg
has been blown away, and the remaining gas is locked up in a few
blobs in the midplane. In contrast, the disks in runs sb-SNCR and sb-SNCR-sg
can survive for a longer time. Therefore, their evolution is more steady compared to the bursty situation in run sb-SN-sg.

The high-resolution run S10-KS-clus behaves similarly as the other simulations but
shows larger fluctuations over time, which is expected. The magnitude of the surface brightness
in the lower-resolution runs is therefore not affected by resolution.

All the simulations reach surface brightness values between $S = 10^{-13}\,$erg cm$^{-2}$ s$^{-1}$ deg$^{-2}$ and
$S = 10^{-11}\,$erg cm$^{-2}$ s$^{-1}$ deg$^{-2}$. This agrees with the observations by \citet{henshe13} within an order of magnitude.
Because the time variability of the observed values is unknown, and since our simulations do not reach a steady state within the simulation runtime,
a comparison of absolute values should not be overemphasized.
The exact surface brightness values depend sensitively on the simulation
parameters, but all of our simulations naturally produce values of the same order of magnitude as the observations.

\section{Energy budget}
\label{sec:ebud}

We can estimate the energy budget of the upper halo ($z \geq 0.1$\,kpc) as follows. For symmetry reasons,
we assume that half of the energy input (thermal or cosmic rays) from supernova explosions
goes into the heating of the upper halo. Furthermore, we have to consider the total X-ray heating rate of the upper
halo due to the \citet{woletal95} prescription, which is the only heating term in our chemical network
that affects hot, ionized gas. Both forms of energy input give the total heating rate $\Gamma_\mathrm{tot}$.
In simulations with the fiducial supernova rate and thermal feedback, the energy input from supernova explosions is about
$10$ times the X-ray heating rate, however in run CR both forms of energy input are comparable in magnitude.
Therefore, the contribution of X-ray heating to the surface brightness is negligible when thermal feedback is present.
The energy radiated away by soft X-rays is given by the luminosity $L_\mathrm{X}$.

Figure~\ref{fig:time} shows $L_\mathrm{X}$ and the ratio $\Gamma_\mathrm{tot} / L_\mathrm{X}$ as function of time. The amount of energy
radiated away in soft X-rays is usually less than a percent and in most cases
even less than a permille.

\section{Conclusions}
\label{sec:con}

We have presented a differential analysis of the soft X-ray emission from the galactic halo in a set of simulations in which we systematically
vary the effects of supernova feedback, cosmic-ray driving and self-gravity.
We find that star formation feedback alone is sufficient to explain the observed surface brightness.
For pure thermal feedback, the results are very sensitive to the supernova rate. 
Cosmic rays suppress this dependence and reduce the surface brightness by an order of magnitude or more.
Self-gravity binds the diffuse gas in the disk plane, so that supernovae exploding in voids can
blow a large amount of hot gas into the halo. This can boost the soft X-ray surface brightness by several orders of magnitude.
Because this quantity is extremely sensitive to the above physical effects, we conclude that observations of the soft X-ray surface brightness
alone do not provide useful diagnostics for the study of galactic star formation.

\section*{Acknowledgements}

We thank Jonathan Mackey for useful discussions and John ZuHone for providing the APEC emissivity file.
We also thank Alex Hill and Mordecai Mac Low for access to their simulation data 
and the referee for helpful comments that improved the clarity of the paper.
T.P., P.G., A.G., T.N., S.W., S.C.O.G., R.S.K., and C.B. acknowledge support from the DFG Priority Program 1573 {\em Physics of the Interstellar Medium}.
T.N. acknowledges support from the DFG cluster of excellence \emph{Origin and Structure of the Universe}.
S.W. acknowledges the support of the Bonn-Cologne Graduate School, which is funded through the Excellence Initiative,
and the SFB~956 \emph{Conditions and impact of star formation} funded by the DFG.
R.W. acknowledges support by the Czech Science Foundation grant 209/12/1795 and by the project RVO:67985815 of the Academy of Sciences of the Czech Republic.
R.S.K., S.C.O.G., and C.B. thank the DFG for funding via the SFB 881 \emph{The Milky Way System} (subprojects B1, B2, and B8). R.S.K. furthermore acknowledges support from the European Research Council under the European Community's Seventh Framework Programme (FP7/2007-2013) via the ERC Advanced Grant STARLIGHT (project number 339177).
The FLASH code was in part developed by the DOE-supported Alliances
Center for Astrophysical Thermonuclear Flashes (ASCI) at the University of Chicago.
The data was analyzed with the {\sc yt} code \citep{turketal11}.

\end{document}